\documentclass[prb,showpacs,preprintnumbers,amsmath,amssymb,preprint,floatfix]{revtex4}

\usepackage{graphicx}
\usepackage{dcolumn}
\usepackage{bm}

\begin{document}

\title{Influence of chopped laser light onto the electronic transport through atomic-sized contacts}

\author {D. C. Guhr, D. Rettinger, J. Boneberg, A. Erbe, P. Leiderer, E. Scheer}
 \affiliation{Fachbereich Physik,
Universit\"at Konstanz, D-78457 Konstanz, Germany}

\date{\today}

\begin{abstract}
\noindent This article reports on the influence of laser
irradiation onto the electrical conductance of gold nanocontacts
established with the mechanically controllable breakjunction
technique (MCB). We concentrate here on the study of reversible
conductance changes which can be as high as 200\%. We investigate
the dependence on the initial conductance of the contacts, the
wavelength, the intensity and position of the laser spot with
respect to the sample. Under most conditions an enhancement of the
conductance is observed. We discuss several physical mechanisms
which might contribute to the observed effect including thermal
expansion, rectification and photon-assisted transport. We
conclude that thermal expansion is not the dominating one.\\
\\
Corresponding author: Elke Scheer (elke.scheer@uni-konstanz.de)

\end{abstract}
 \maketitle

\section{Introduction}
\noindent Metallic nanostructures with fine tips and sharp edges
are interesting candidates for optical antennae due to possible
field enhancement (FE) effects at the parts with small curvatures
and narrow gaps. The excitation of resonant plasmons can amplify
the electrical field at positions within or close to the
nanostructures. The field enhancement in bow-tie shaped metal
structures has so far been calculated and detected by several near
field optical methods \cite{hecht} and direct imaging via ablation
of the underlying substrate \cite{leidererapl}. Here we address
the question whether FE can be detected in the electronic
transport between two metallic tips connected via an atomic-size
contact. The central part of the device resembles a bow-tie
\cite{bowtie}. For this geometry to our knowledge no calculations
of the magnitude of the FE do exist, however, it can be assumed
that as long as the coupling between the two electrodes is weak,
it should still exist. Such contacts can be realized via
mechanically controllable breakjunctions (MCBs), i.e.
free-standing nanobridges with lateral sizes in the order of 100
nm and bridge lengths of several micrometers, which have been
fabricated on a flexible substrate. By bending the substrate the
bridge is elongated and its cross-section is reduced until it
finally breaks at some position. Just before it breaks the
constriction forms a contact with one or a few atoms in cross
section.

These contacts represent model systems which are regularly used
for revealing and understanding the electronic transport
properties of atomic scale circuits \cite{agrait03}. In this limit
the electronic transport can be attributed to a small number of
independent electronic modes, nick-named "conduction channels"
\cite{landauer70}. The transport properties are fully described by
a set $\left\{ \tau_{\mathrm i}\right\}
=\left\{\tau_1,\tau_2,...\tau_{\mathrm N}\right\} $ of
transmission coefficients which depends both on the chemical
properties of the atoms forming the contact and on their
geometrical arrangement \cite{aluprl,nature}. The conductance of
such a contact is given by the Landauer formula:
$G~=~G_0\thinspace \sum\limits_{i=1}^N \tau_{\mathrm i}$, where
$G_0=2e^2/h$ is the conductance quantum.

Besides this aspect, MCBs can be broken to form electrode pairs
with a narrow gap in the order of picometers to nanometers between
two very fine metallic tips. Such electrodes are used for
contacting individual nanoobjects such as clusters or molecules
\cite{springer_molelect,reichert,boehler}. Here we address the
question whether the MCB electrodes are suitable optical antennae
as well. Contrary to planar electrode pairs on a substrate, the
MCB electrodes provide the possibility to rearrange the geometry
of the device and therewith the FE effects in the constriction. By
breaking the nanobridge to separate electrodes one reduces the
interaction between the electrodes, but for small distances it is
still possible to measure the tunneling current. In particular, in
the so-called contact regime the interaction between the tips has
to be taken into account for understanding the optical properties.
The electronic transport is in the quantum regime described above.

 Due to the
electronic structure of molecules with well defined gaps between
the individual molecular orbitals the current-voltage
characteristics (IVs) of single-molecule devices are strongly
nonlinear \cite{reichert}. Under certain conditions the
irradiation with laser light can be regarded as the addition of an
ac-electrical field to the dc voltage used for the transport
measurement. In this situation rectification of the current and
the creation of asymmetries in the IVs is expected \cite{moeller}.
Furthermore, by irradiation with laser light with a suitable
frequency photochromic molecules may undergo conformational
changes which may be reflected in the electronic transport
properties. Duli\'{c} et al. observed a switching from the "on" ,
i.e. low resistance state of a molecular contact to the "off"
state by shining light with a frequency which is known to create a
conformation change when the molecules are in solution or
deposited in an ensemble on a surface \cite{dulic}. However, the
switching from "off" to "on" was not observed. The interpretation
of those resistance changes is hampered by the fact that several
explanations are possible for the observation of a strong
resistance increase, among which the breaking of the contact due
to geometry of the electrodes is one of the most likely scenarios.

In our work we therefore concentrate on the study of laser
illumination onto the metal electrodes alone. Already for this
conceptually rather simple device the influence of the laser light
may be manifold. One rather trivial effect is the geometry change
of the tips due to thermal expansion because of the deposited
energy of the laser pulse. This effect has been shown to be the
dominating one for a scanning tunneling microscope (STM) under
pulsed laser light irradiation \cite{bonebergSTM,grafstroem}.
However, for the MCB geometry this effect is expected to be much
smaller due to the very small dimensions of the  freestanding
bridge arms.

The electromagnetic wave of the laser pulse represents a
high-frequency electrical field that is coupled to the metal
bridge. The ac field may create an asymmetric contribution to the
dc current for those voltages where the IVs are nonlinear
\cite{moeller}. Asymmetric irradiation of the contact might create
temperature gradients  - and thus thermo-currents - ,  or
photocurrents. Last but not least, photon assisted transport
(PAT), i.e. the creation of quasiparticles with an energy $\hbar
\omega$ above the Fermi energy will create a nonequilibrium in the
electronic system. This means that high-energetic quasiparticles
may contribute to the transport. Due to the energy dependence of
the transmission coefficients of the conduction channels the
resistance of the contact will change \cite{viljas}.

We will discuss these mechanisms in the light of our experimental
results and come to the conclusion that besides thermal expansion
at least one additional mechanism is active.

\noindent\section{Experimental}

\subsection{MCB technique and fabrication of
lithographic breakjunctions}

\begin{figure}
 \begin{center}
\includegraphics[width=0.85\columnwidth,clip]{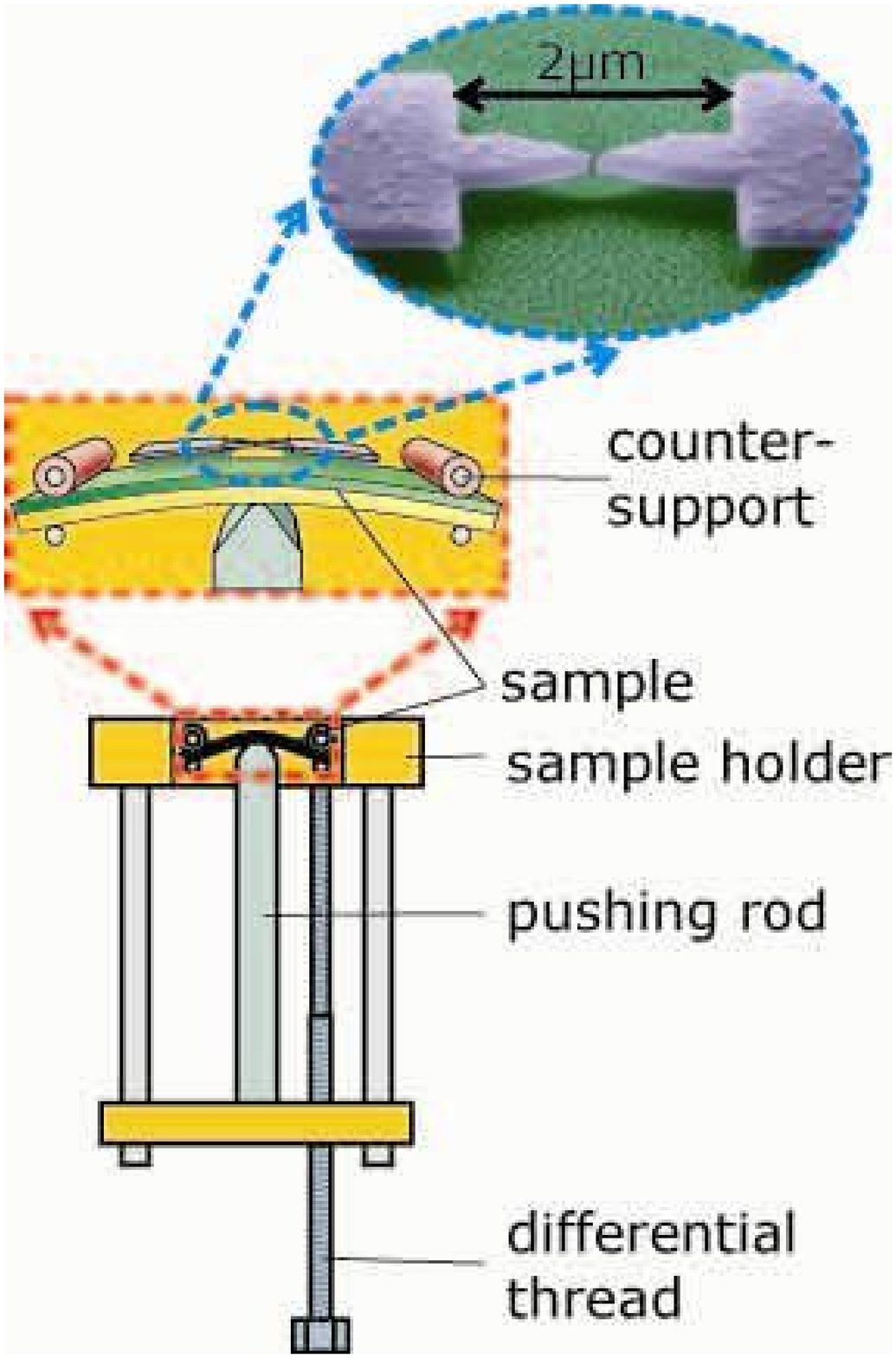}
    \caption{\label{guhr_f1} a) Scheme of the mechanically controllable breakjunction setup (sample dimensions not to scale) and
electron micrograph of a typical sample.}
\end{center}
\end{figure}

\noindent The nanoelectrodes are arranged by the mechanically
controllable breakjunction (MCB) technique \cite{agrait03}. MCBs
consist of a freestanding micro- or nanobridge on a flexible
substrate embedded in an electrical circuit by leads which are in
rigid contact with the substrate (see Fig. \ref{guhr_f1}). By
bending the substrate with the help of a mechanical or piezo drive
the freestanding parts are elongated and constricted. The
dimensions can be chosen such that constrictions with lateral
dimensions of a single or few atoms or vacuum tunnel contacts
between two atomically sharp tips can be arranged and stabilized.
The main advantage of the MCB technique compared to atomic size
contacts arranged with STMs is the built-in stability due to the
high reduction ratio $r$ (between the motion of the pushing rod
and the displacement of the bridge arms). Mechanical vibrations
are damped with a factor $1/r$ which enables MCBs to be used
without external vibration damping even when mounted in vacuum
chambers with mechanical pumps. Different realizations of this
technique have been developed, among which the lithographic MCBs
that we apply here offer the possibility to use a purely
mechanical drive and to incorporate the device into more complex
electronic circuits.

 The samples are
fabricated by electron beam lithography along the lines of Ref.
\onlinecite{jansaclay}. The suspended nanobridges are 2\thinspace
$\mu $m long and 100 nm
thick made of Au with a $200$\thinspace ${\rm nm}\times 100$\thinspace ${\rm %
nm}$ constriction in the center (cf. insert of
Fig.~\ref{guhr_f1}). The thickness of the flexible steel
substrates is $\approx 0.3\thinspace$mm, the distance between the
counter-supports of the bending mechanism is 19~mm.  The metallic
substrate is covered with a 2\thinspace$\mu$m thick polyimide
layer which serves for planarization, electrical isolation and as
sacrificial layer for suspending the central part of the sample.
The bridge is underetched in an isotropic oxygen plasma which
reduces the height of the polyimide layer by about 500 to 1000~nm.
At the narrow constriction the nanobridge is now unsupported. The
resistance of the nanobridge does not increase during the etching
process, indicating that the reaction between the oxygen ions and
Au is weak.

The samples are then mounted on a three-point bending mechanism as
shown in Fig. \ref{guhr_f1}. A differential screw with
100\thinspace $\mu $m pitch, driven by a dc-motor and an
(exchangeable) reduction gear box, controls the motion of the
pushing rod that bends the substrate. With these sample parameters
we achieve reduction ratios of the order of 30000:1. The relative
displacement of the pushing rod can be controlled to a precision
of $\approx \thinspace 30\thinspace $nm, which thus results in a
relative motion of the two anchor points of the bridge of around
1\thinspace pm. This was verified using the exponential dependence
of the conductance on the inter-electrode distance in the tunnel
regime.

Typical opening speeds for the continuous measurements (see below)
are 215 nm/s for the pushing rod, corresponding to a displacement
of the bridge-anchor points of 7 pm/s, much slower than typical
opening speeds of nanocontacts fabricated with the help of a STM
or the "notched-wire" breakjunction technique \cite{agrait03}.
This limitation in opening speed, however, limits the number of
opening traces that can be recorded within a reasonable time.

The mechanics are designed such that the position of the
nanobridge remains constant upon bending the substrate in order to
ensure stable optical focal conditions (see below). The MCB setup
is operational at ambient conditions as well as in vacuum and or
at low temperatures. The measurements presented in this article
have been performed under UHV conditions at room temperature on
three nominally identical samples.

\subsection{Optical setup} 
\begin{figure}
  \begin{center}
\includegraphics[width=0.95\columnwidth,clip]{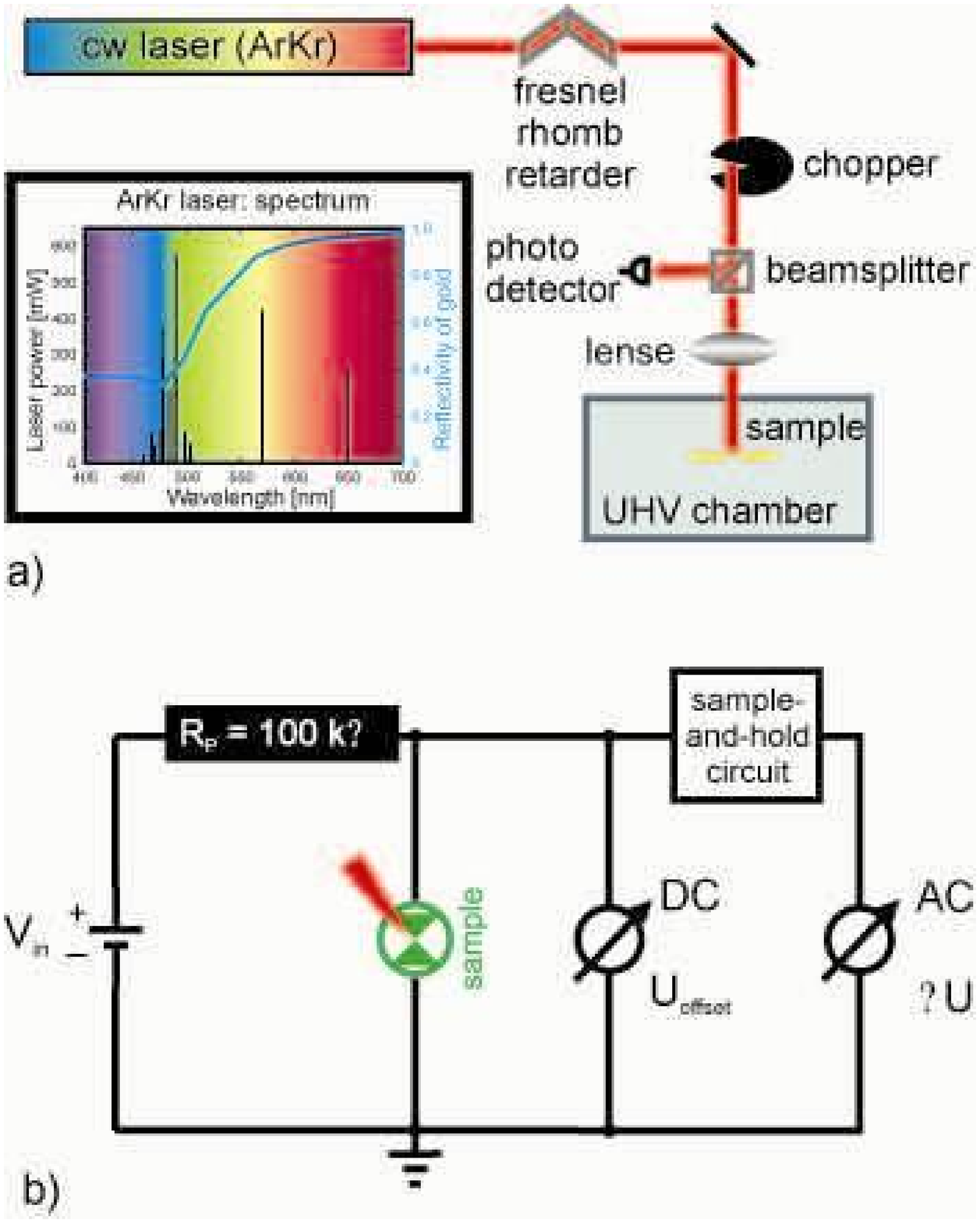}
    \caption{\label{guhr_f2} a) Scheme of the optical setup of the experiment. The inset shows the intensity spectrum of the ArKr laser
    and the reflectivity of gold \protect\cite{HandChemPhys82}. b) Scheme
    of the electronic measurement circuit.}
    \end{center}
\end{figure}

\noindent
The optical setup is depicted in Fig. \ref{guhr_f2} a. We use an
Argon-Krypton continuous wave (cw) laser system as light source,
which allows us to select one of several wavelengths in the range
between 480\thinspace nm and 650\thinspace nm. The cw laser beam
is fed through a $\frac{\lambda}{2}$\thinspace-\thinspace Fresnel
rhomb retarder in order to tilt the polarization angle, and is
afterwards chopped to produce light pulses with a length of
700\thinspace $\mu$s and a repetition rate of 50\thinspace Hz. A
beam-splitter in the optical path divides the chopped light. One
part is reflected into a photo detector to create a trigger event,
the other part is focussed onto the sample by a lens. The beam
spot on the sample has a diameter of about 20 to
25$\thinspace\mu$m, depending on the wavelength.

The sample is mounted on the breaking mechanism inside a UHV
chamber with a base pressure of 10$^{-9}$\thinspace mbar in order
to avoid contamination of the breakjunction. The UHV chamber is
equipped with a fused silica-window to minimize the change of
polarization of the laser light due to stress birefringence. If
not stated differently, the polarization was chosen parallel to
the current for the measurements presented here. For perpendicular
polarization similar results but with an amplitude that is smaller
by about a factor 2 to 3 are observed. A more detailed
investigation is necessary in order to quantify the exact
influence of the polarization.

\subsection{Electronic circuit} \noindent
The sample is connected with a 100\thinspace k$\Omega$ series
resistor to a dc-voltage source (Fig. \ref{guhr_f2} b). This
resistor is required to protect the sample against high-voltage
spikes created e.g. by electrostatic discharges. The combination
with the sample builds a voltage divider. Thus, we have the
possibility to calculate the conductance of the sample from the
measured voltage across it.

\begin{figure}
  \begin{center}
\includegraphics[width=1.00\columnwidth,clip]{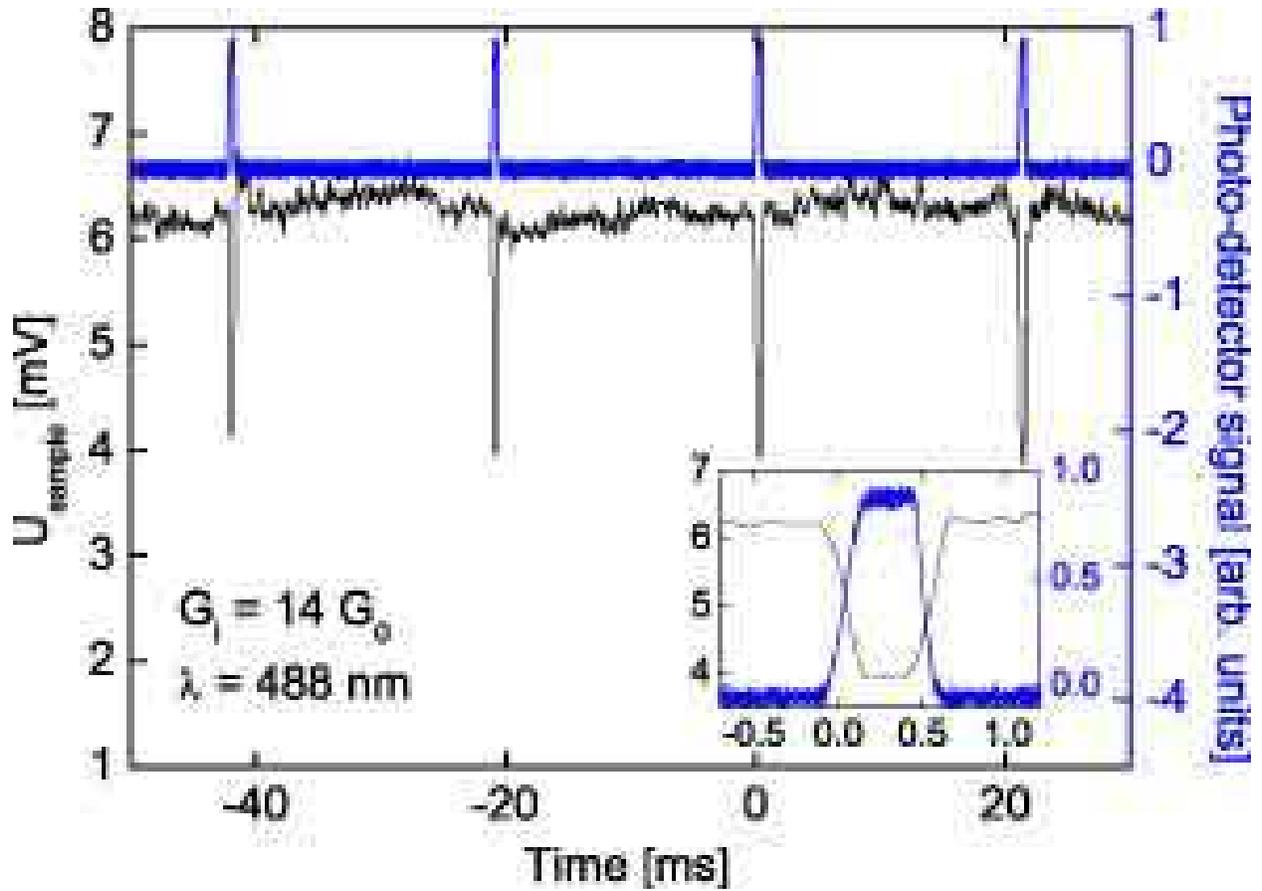}
    \caption{\label{guhr_f3} Signal of the photo detector (blue) and the integrating
    amplifier (black)
     as a function of time showing four consecutive light pulses.
    The inset shows both signals for one pulse on a magnified scale, visualizing
    the time resolution of the measurement limited by the opening speed of the chopper. The voltage
     signal follows the finite slope of the photo detector signal, indicating roughly linear intensity dependence of the
     LIS}.
    \end{center}
\end{figure}

Because of the different voltage ranges of the dc-voltage signal
(without light) and the small voltage variation due to the
light-induced-signal (LIS$=\Delta G/G$) during the pulse, it is
necessary to measure the signal across the sample in dc- as well
as in ac-mode in order to get a good resolution in both parts. The
amplitude of the ac-part, caused by each light pulse, is measured
by a sample-and-hold circuit (S/H) with an integrating amplifier.
The acquired amplitude is held and displayed over the whole
chopping period, until the next light pulse appears. Thereafter
the next voltage variation is measured by the (S/H)
circuit and held again.\\
To calculate the conductance variation caused by the incident
light, we convert the dc-voltage and the sum of the dc- and
ac-signal (accounting for the amplification of the (S/H) circuit)
into conductance and compare the values. The time resolution of
the experiment is limited by the fact that the chopper is located
at a position where the laser spot is rather large. Thus, the
laser intensity rises and drops on a time scale of 100~$\mu$s. The
response time of LIS is similar demonstrating a roughly linear
intensity dependence of the LIS (see inset of Fig.~\ref{guhr_f3}).

\subsection{Control experiments}
\begin{figure}
  \begin{center}
\includegraphics[width=0.95\columnwidth,clip]{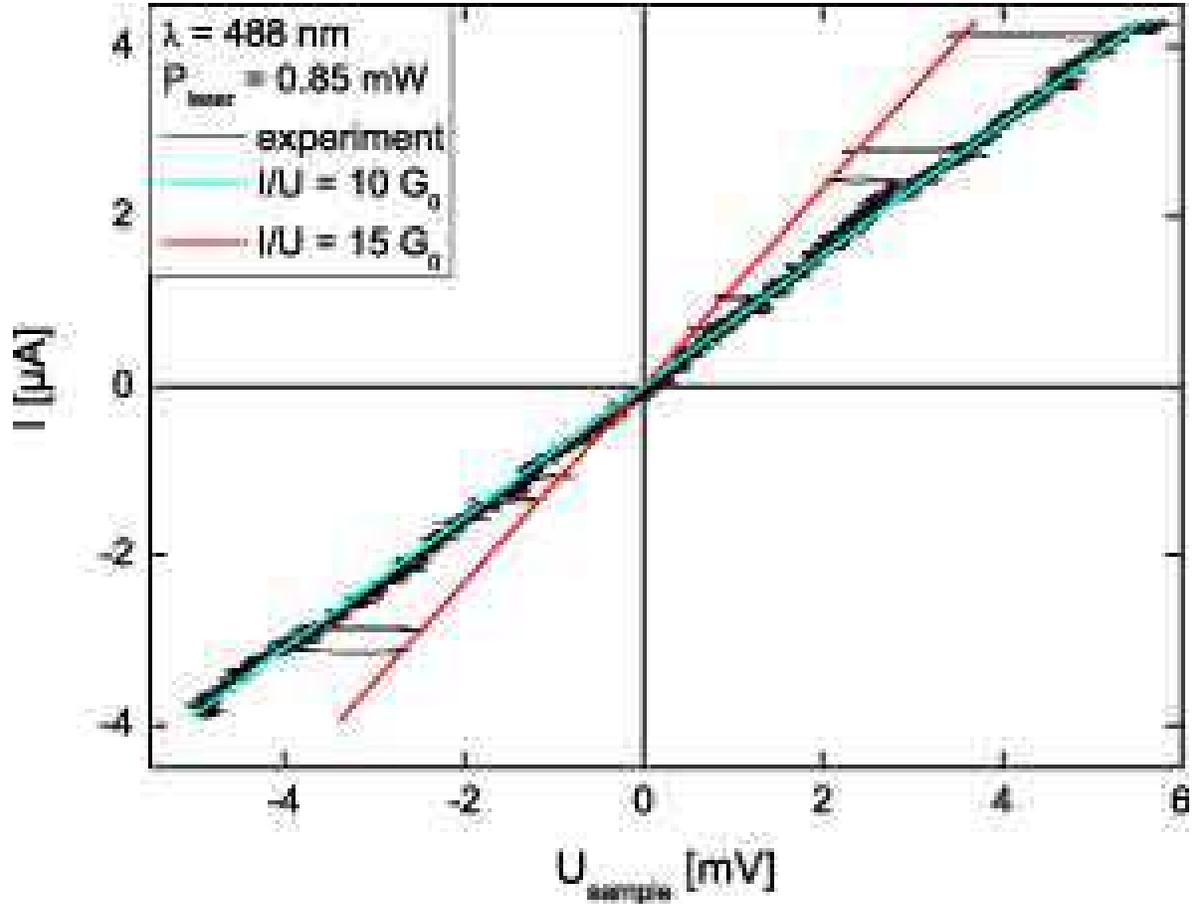}
    \caption{\label{guhr_f4} Current-voltage (IV) characteristic of a contact with $G_i = 10 G_0$
     showing several light pulses with $P = 0.85~$mW and $\lambda = 488~$nm. During the light
     pulses the conductance is enhanced as indicated by the guide to the eye showing an IV for $G = 15 G_0$.
     The current has been determined by the voltage across the series resistor (see Fig.~\protect\ref{guhr_f2}).}
    \end{center}
\end{figure}

\noindent Figure \ref{guhr_f4} shows a current-voltage
characteristic (IV) of a contact for a conductance without
irradiation of $G_i = 10 G_0$, as indicated by the blue line. At
several current values laser pulses with duration $\tau = 700
\mu$s, $\lambda = 488~$nm and power $P = 0.85~$mW are shed onto
the contact. During the pulse the absolute value of the voltage
measured across the contact is reduced by the same amount for both
polarities. The amplitude of the laser induced voltage drop is
proportional to the current. Neither current nor voltage offsets
nor asymmetries are observed, indicating that the effect of the
light is a bare change of conductance to $G_f = 15 G_0$ as
indicated by the red line. The fact that the conductance change is
observed for linear IVs indicates that it is not created by a
rectification effect as observed in Ref. \onlinecite{moeller}.

 The
conductance change is completely reversible and reproducible for a
given contact as demonstrated by shining repeated pulses onto the
bridge (see Fig. \ref{guhr_f3}); $G$ only changes upon variation
of the contact geometry. However, continuous irradiation of the
device with $\lambda = 488\thinspace$nm for several seconds with a
similar power ($\approx$ mW) results in irreversible conductance
changes (not shown), indicating that irreversible atomic
rearrangements are triggered by the cw light.

\section{Results and discussion}

\noindent When elongating the bridge without irradiation, its
conductance $G$ decreases in steps of the order of 1\thinspace
$G_0$, their exact sequence changing from opening to opening
because the atomic arrangement of the central region differs for
each opening process. The typical distance scale for a conductance
change of $1~G_0$ is the lattice constant, i.e. a few tenths of a
nm. Because of the individual rearrangements it is difficult to
draw conclusions from the behavior of individual contacts or
openings. Experiments on a large ensemble of metallic contacts
have demonstrated the {\it statistical} tendency of atomic-size
contacts to adopt element-specific preferred values of
conductance. The actual preferred values depend not only on the
metal under investigation but also on the experimental conditions.
However, for many metals, and in particular monovalent metals like
Au, the smallest contacts have a conductance $G$ close to $G_0$
(Ref. \onlinecite{agrait03}).

\begin{figure}
  \begin{center}
\includegraphics[width=0.9\columnwidth,clip]{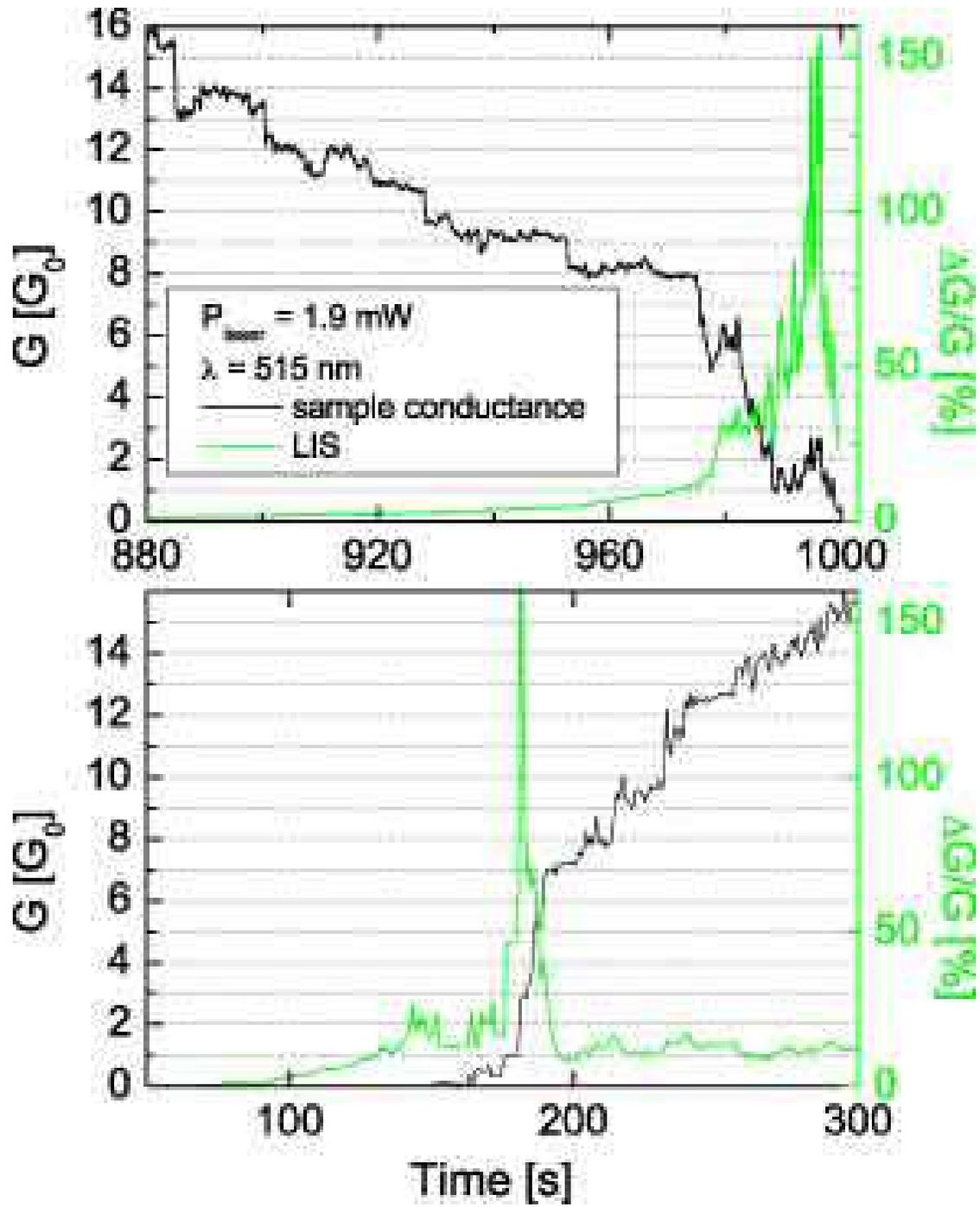}
    \caption{\label{guhr_f5} Upper (lower) panel: Conductance and light-induced relative conductance change $\Delta G/G$ as a
    function of time when opening (closing) the breakjunction continuously.
    The wavelength was $\lambda = 515~$nm and the laser power was $P = 1.9$~mW with polarization parallel to the current direction.}
    \end{center}
\end{figure}

Also the LIS varies from contact to contact. Therefore, in order
to deduce the typical behavior, we simultaneously measure the
conductance and the LIS upon continuously opening and closing the
bridge. Fig.~\ref{guhr_f5} gives an example of such measurements
recorded for green light with $\lambda = 515$~nm. The conductance
plateaus are not always very well marked. The tunneling regime is
difficult to assess when opening because the contact tip atoms
relax back upon breaking giving rise to strong decrease in the
conductance by several
orders of magnitude \cite{agrait03}. 
When closing the contact again, very often the first contact after
the "jump to contact" has already a conductance of several $G_0$.
The LIS is positive throughout the whole opening process. In some
occasions it adopts small negative values in the tunneling regime
$G \ll G_0$ (not shown in this example). It reaches a maximum with
large fluctuations for $G \simeq 2 G_0$. From these observations
we shall draw the important conclusion that thermal expansion
alone cannot explain the LIS. In similar experiments but using an
STM instead of a MCB, it has been shown that thermal expansion is
the most important mechanism giving rise to conductance changes
\cite{bonebergSTM,grafstroem}.

 A priori it is not predictable whether
in our geometry thermal expansion should give rise to an
enhancement or a reduction of the conductance, since the laser
spot also hits the underlying substrate. If the thermal expansion
of the nanobridge is larger than the one of the substrate, the two
electrodes are pushed together, if the thermal expansion of the
substrate exceeds the one of the nanobridge, then the two tips are
pulled apart from each other. Whether the motion of the tips
results in an enhancement or reduction of $G$ depends on the
plateau shape of the conductance trace upon opening: For a
horizontal plateau the conductance is constant upon stretching the
contact, i.e. changing the bridge length does not give any
conductance change. For negatively declined plateaus, i.e.
decreasing $G$ upon stretching, an elongation of the tips yields
an enhancement of $G$ while positively declined plateaus give rise
to a decreasing $G$ when pushing the tips together. The
conductance trace of Fig.~\ref{guhr_f5} shows horizontal as well
as negatively and (only rarely) positively declined plateaus.
Nevertheless, in our experiments the LIS is almost always found to
be positive (see below). Furthermore, the strong exponential
distance dependence of $G$ for vacuum tunneling should give rise
to a large relative conductance change upon irradiation. However,
our observed LIS is usually smallest in this regime. In
Fig.~\ref{guhr_f5} we chose an example where it is particularly
large and of comparable size with the signal for large contacts.
Hence, although thermal expansion is very likely to be present in
our experiment, it is expected not to be the dominating effect for
the LIS.

\begin{figure}
  \begin{center}
\includegraphics[width=1.00\columnwidth,clip]{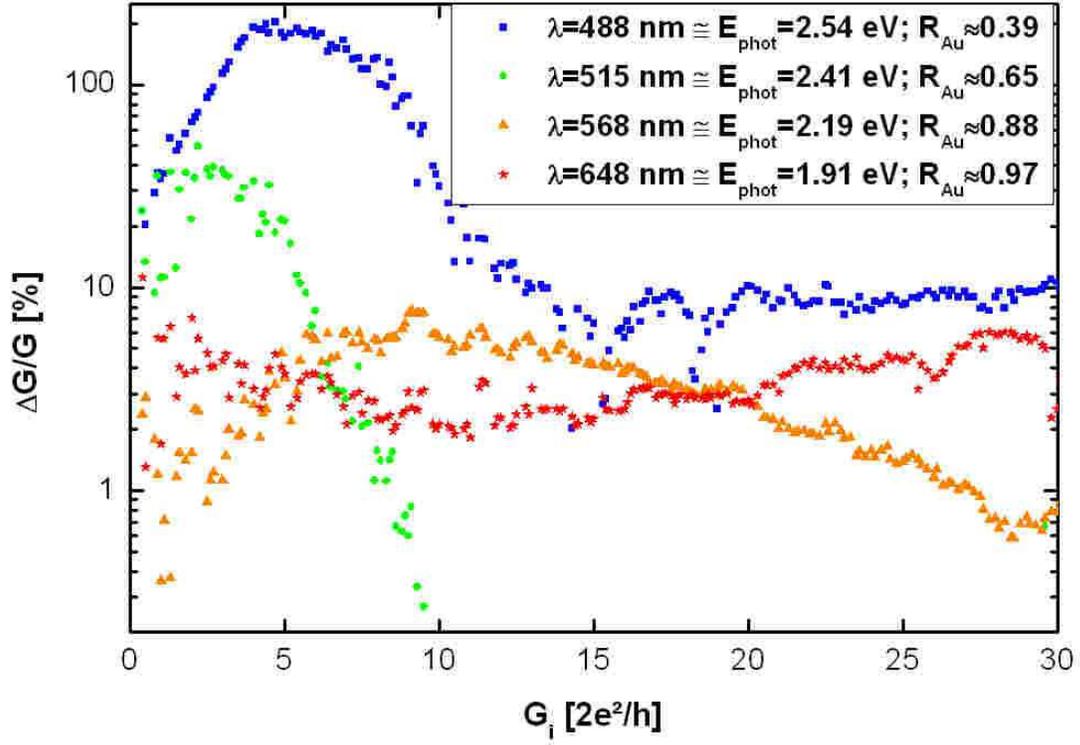}
    \caption{\label{guhr_f6} Histogram of of the light-induced signal
    $\Delta G/G$ vs. $G_i$ for different laser wavelengths $\lambda$. The amplitude
    as well as the position of the maximum strongly depend on $\lambda$. $E_{phot}$ is the energy per photon and $R_{Au}$ is the reflection coefficient of electropolished Gold (110)\cite{HandChemPhys82}.}
    \end{center}
\end{figure}

In order to further elucidate the origin of the effect, we
performed statistical investigations of the LIS for different
light wavelengths. Fig.~\ref{guhr_f6} depicts histograms of the
relative conductance changes recorded on the same sample with
similar laser power for four wavelengths and from analyzing
between 35 and 170 opening traces each. The histograms are
calculated as follows: For each opening trace the LIS is related
to the initial conductance value $G_i$ without laser irradiation
for which it was measured. The average LIS and its standard
deviation are then plotted on a logarithmic scale versus $G_i$.

As seen from Fig.~\ref{guhr_f6} the signal is largest for blue
light with $\lambda = 488$~nm and decreases with increasing
wavelength. The LIS also depends on the conductance: it is largest
for few-atom contacts and decreases for smaller and larger
contacts. The $G$ value for which the maximum signal is observed
does not depend systematically on $\lambda$. Although for certain
opening traces the LIS is negative in the tunnel regime, the
average values plotted here are positive.

The amplitude dependence of the LIS on $\lambda$ correlates with
the variation of the reflectivity of gold with the wavelength,
which increases from 39\% for the blue light up to 97\% for the
red light. This means that the deposited energy decreases by a
factor of $\approx 20$, while the maximum signal size decreases by
about a factor of 40. The additional suppression of the LIS in the
red is in accordance with the predictions of photo assisted
transport \cite{viljas} (see below.)

\begin{figure}
  \begin{center}
\includegraphics[width=1.0\columnwidth,clip]{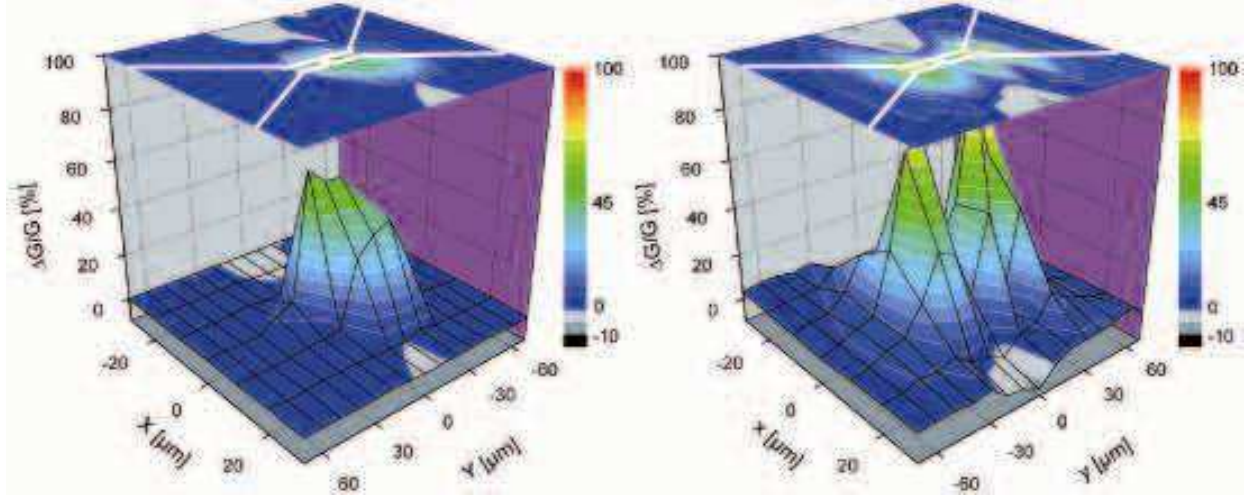}
    \caption{\label{guhr_f7} Position map of $\Delta G/G$ for a (b)
    $\lambda = 488~$nm (648~nm) for $G_i = 6.5~ G_0$, $P = 2.2~$mW (23~mW) and polarization
    perpendicular to the current.
    The signal has been determined by averaging over of 20 (10) consecutive light pulses for each position. The spatial stepsize is $10~\mu$m.
     As a guide to the eye we show the contour of the sample.}
    \end{center}
\end{figure}

Further information about the effect is obtained from recording a
map of the signal size with respect to the position of the laser
spot. The maps for blue and red light for an average conductance
of $G = 6.5~G_0$ and polarization perpendicular to the transport
direction are shown in Fig.~\ref{guhr_f7}. The laser spot with a
diameter of $\approx 20~\mu$m is much larger than the bridge
length, therefore the step size for the map is set to $10~\mu$m. A
clear position dependence is observable for all wavelengths. In
Fig.~\ref{guhr_f7} we superimposed a drawing of the sample
geometry with the recorded LIS. The largest signal is observed
when the center of the laser spot is located on the leads within a
distance of $20~\mu$m from the contact. It is symmetrical with
respect to the contact. For red light a relative minimum is
observed right at the position of the contact, while for blue
light the signal is almost constant along the bridge. More than a
diameter of the laser spot aside of the contact a small negative
LIS is observed for the red light.

These observations can tentatively be explained as follows: The
negative signal when mainly irradiating the substrate indicates
that the effect of thermal expansion of the substrate tends to
decrease the conductance. Thus, the observed enhancement when the
laser spot does touch the metallic bridge, is due to light coupled
into the metallic bridge.
 When irradiating the narrow constriction, only a small part of the
 laser energy is coupled into the metallic bridge, but the
 majority is absorbed by the substrate. Thus the minimum in the
 LIS can be explained by the geometry of the experiment. Contrary,
 the lack of minimum for the blue light although the geometry of the experiment is the same
 indicates that an energy dependent mechanism is at the origin of
 the strong enhancement.\\

 A possible explanation could be photon assisted transport
 reflecting the energy dependence of the transmission coefficients
 of the conduction channels as proposed by Viljas and Cuevas
 \cite{viljas}. The mechanism is such that the incoming photons are
 absorbed by individual electrons, creating quasiparticles with energy $\hbar \omega$
 below and above the respective Fermi energies of both electrodes. Since electrons on both sides of the constriction are
 concerned,
 the quasiparticles may travel in both directions. The
 transmission coefficients of the conduction channels reflect the
 local density of states at the central atom. Although the exact functional
 dependence of the transmission differs from that of the density of states, the typical energy scales are identical.
 For single-atom contacts of Au it has been predicted that the typical energy scale for non-negligible variation of the
 transmission coefficients is about one to several electron volts
 \cite{cuevasplateaus}.
Those quasiparticles have a lifetime in the order of several ten
femtoseconds, depending on their energy \cite{lifetimes}. It has
been shown that the effective thermalization times are much longer
than the lifetimes estimated from Fermi liquid theory
\cite{guo,aeschlimann,lifetimes}. Since furthermore the transport
mode of the highly excited quasiparticles differs from the thermal
ones close to the Fermi edge, it is difficult to determine a
corresponding range or scattering length \cite{lifetimes}. A
rather optimistic estimate yields a length in the order of a few
hundred nanometers. This means that only incoming photons within
this distance from the contact are able to create the
high-energetic quasiparticles which give rise to the conductance
change. For photon energies of a few eV, strong
 variations of the conductance are thus expected, the sign of
 which is not easy to predict because of the particular energy landscape of gold atomic-size contacts
 \cite{viljaspriv}. The dominating transmission of the s-band
 decreases for smaller as well as for higher energies, but for lower
 energies the 5d-bands and for higher energies the 6p-bands start
 to contribute to the transport. Since no detailed calculation for
 gold does exist by now, a quantitative analysis in terms of PAT is
 not possible yet.

Because of the short lifetime of the high energetic quasiparticles
one would thus expect to have a maximal signal closest to the
contact, but because of the reduced lateral size less photons
arrive at the metal bridge. For blue light the increase of the
efficiency onto the transport and the loss in number of photons
seem to compensate while for red photons the loss in number
dominates. We speculate that the influence of the red light with
photon energy below the plasma edge and blue light above it might
be of different nature. Below $\hbar \omega_p \simeq 2.5~$eV (see
Ref.~\onlinecite{simmons}) corresponding to $\lambda \approx
500~$nm a hot electron sea might be created which distributes the
energy among the plasma electrons, while above $\hbar \omega_p$
the energy transfer to individual electrons is more likely.
Furthermore, the strong energy dependence of the lifetime yields
considerably shorter lifetimes for blue than for red light.
Further experiments are necessary to verify this interpretation.
 Nevertheless we would like to mention that the
 observed wavelength dependence and amplitude of the signal are in
 accordance with this mechanism.

Taking into account the geometry of the sample, the focal
cross-section, the intensity of the laser, pulse length, and the
reflectivity at a given wavelength we obtain a rough estimate for
the number of created high-energy quasiparticles per photon. For
$\lambda = 488$~nm and the contact shown in Fig. \ref{guhr_f3} at
an applied voltage of 1~mV this value is $QE = 10^{-13}$.

\section{Summary}
\noindent We have presented electric transport measurements
carried out on atomic-size contacts connecting the two parts of a
bow-tie shaped optical-antenna geometry under irradiation with
laser light. We observe strong conductance enhancements, the
amplitude of which strongly depends on the wavelength of the
laser. We discuss several possible mechanisms and come to the
conclusion that besides thermal expansion photon-assisted
transport is most likely at the origin of our observations.

\section{Acknowledgements}
\noindent   Fruitful discussions with J.C. Cuevas, J.K. Viljas and
A. Leitenstorfer are gratefully acknowledged. This work was
financially supported by the Deutsche Forschungsgemeinschaft
through SFB 513 and the Alfried Krupp von Bohlen und
Halbach-Stiftung.


\newpage

\newpage

\newpage

\newpage

\newpage

\newpage
\end{document}